\documentclass[aps,prb,reprint,groupedaddress,footinbib,citeautoscript]{revtex4-1}

\usepackage{graphicx}
\graphicspath{{.}{./EPS/}}

\usepackage[utf8]{inputenc}
\usepackage{csquotes}
\usepackage[american]{babel}
\usepackage[T1]{fontenc}
\usepackage{enumerate}
\usepackage{mdwlist}
\usepackage[activate=normal]{pdfcprot}
\usepackage{bbding}
\usepackage{color}
\usepackage{amssymb}
\usepackage{amsmath}
\usepackage{amsfonts}
\usepackage{mathrsfs}

\usepackage{bm}
\usepackage{dcolumn}
\usepackage{color}
\usepackage[colorlinks=true,citecolor=blue]{hyperref}
\hypersetup{colorlinks=true,citecolor=blue,linkcolor=red,urlcolor=blue}

\newcommand* {\vek}[1]{{\ensuremath{\bm{\mathrm{#1}}}}}

\newcommand* {\kk}{\vek{k}}
\newcommand* {\rr}{\vek{r}}

\newcommand* {\ket}[1]{\ensuremath{| {#1} \rangle}}

\newcommand* {\ham}{\mathsf{H}}
\newcommand*{\ee}{\ensuremath{\mathrm{e}}}

\begin{document}

\title{Exporting superconductivity across the gap: Proximity effect for semiconductor
valence-band states due to contact with a simple-metal superconductor}

\author{A.~G. Moghaddam}
\affiliation{Department of Physics, Institute for Advanced Studies in Basic
Sciences (IASBS), Zanjan 45137-66731, Iran}

\author{T. Kernreiter}
\affiliation{School of Chemical and Physical Sciences and MacDiarmid Institute
for Advanced Materials and Nanotechnology, Victoria University of Wellington,
PO Box 600, Wellington 6140, New Zealand}

\author{M. Governale}
\affiliation{School of Chemical and Physical Sciences and MacDiarmid Institute
for Advanced Materials and Nanotechnology, Victoria University of Wellington,
PO Box 600, Wellington 6140, New Zealand}

\author{U. Z\"ulicke}
\email{uli.zuelicke@vuw.ac.nz}
\affiliation{School of Chemical and Physical Sciences and MacDiarmid Institute
for Advanced Materials and Nanotechnology, Victoria University of Wellington,
PO Box 600, Wellington 6140, New Zealand}

\date{\today}

\begin{abstract}
The proximity effect refers to the phenomenon whereby superconducting properties
are induced in a normal conductor that is in contact with an intrinsically superconducting
material. In particular, the combination of nano-structured semiconductors with
bulk superconductors is of interest because these systems can host unconventional
electronic excitations such as Majorana fermions when the semiconductor's charge
carriers are subject to a large spin-orbit coupling. The latter requirement generally
favors the use of hole-doped semiconductors. On the other hand, basic symmetry
considerations imply that states from typical simple-metal superconductors will
predominantly couple to a semiconductor's conduction-band states and, therefore,
in the first instance generate a proximity effect for band electrons rather than holes.
In this article, we show how the superconducting correlations in the conduction band
are transferred also to hole states in the valence band by virtue of inter-band coupling.
A general theory of the superconducting proximity effect for bulk and low-dimensional
hole systems is presented. The interplay of inter-band coupling and quantum confinement
is found to result in unusual wave-vector dependencies of the induced superconducting
gap parameters. One particularly appealing consequence is the density tunability of
the proximity effect in hole quantum wells and nanowires, which creates new
possibilities for manipulating the transition to nontrivial topological phases in
these systems.
\end{abstract}

\maketitle

\section{Introduction \& Motivation}

Superconductor-semiconductor heterostructures have been the subject of intense
study~\cite{hekk94,bee95,van97,lam98,sch01,tak08} because they offer intriguing
possibilities to observe effects of quantum phase coherence in electronic
transport~\cite{akk95,soh97,bee97}. The contact to a superconducting material
induces pair correlations of charge carriers in the semiconductor and, especially in
low-dimensional or nanostructured systems, results in a gapped spectrum of
electronic excitations~\cite{deg64,vol95,fag05,kop11}. Recent work has focused
on the interplay of proximity-induced superconductivity and strong spin-orbit coupling
in nanowires~\cite{lut10,ore10,sau10,ali10,kli13,zyu13}, which can give rise to the
presence of unconventional quasiparticle excitations~\cite{ali12,lei12,bee13,sta13}.
As the charge carriers from the valence band of typical semiconductors ("holes") are
generically subject to particularly strong spin-orbit-coupling effects~\cite{win03,klo11},
it has been suggested~\cite{mao11,mao12} that the use of hole-doped nanowires
is a good strategy for experimental realization and detailed study of exotic
quasiparticles. These developments have created a need for a fundamental, and
experimentally relevant, theoretical description of the proximity effect for holes,
which we are providing in this work.

We consider heterostructures consisting of a bulk superconductor in contact with
semiconductors of varying dimensionality. The superconducting material is assumed
to be a simple metal, hence its unfilled band has \textit{s}-like character and couples
only to the semiconductor's conduction-band states because these have compatible
symmetry properties. (States from a typical semiconductor's valence band have
\textit{p}-like symmetry~\cite{yu10}.) The resulting proximity effect can thus induce
a gap only in the dispersion of charge carriers from the conduction band, leaving
the valence band initially unaffected. However, as we show in greater detail below,
the proximity-induced change in the electronic properties of the conduction band
affects also valence-band states via the ubiquitous interband
coupling~\cite{kan57,win03,yu10} that is present at finite wave vector $\kk$.
Previous work~\cite{fut11} has investigated how Andreev reflection of holes is
enabled, and its characteristics changed from that occurring at ordinary
superconductor--normal-metal interfaces~\cite{and64,deg63}, by interband
coupling. Here we generalize this concept to develop a fundamental study of how
superconducting correlations translate from the conduction band into the valence
band. In particular, the method of L\"owdin partitioning~\cite{win03} is employed
to derive the effective Bogoliubov-de~Gennes (BdG) Hamiltonian~\cite{deg89}
that governs the proximity effect for holes, thus providing a starting point for 
further detailed studies of topological phases in hole-doped semiconductor
nanostructures~\cite{mao11,mao12}.

The remainder of this article is organized as follows. Basics of the mathematical
formalism are presented in Sec.~\ref{sec:basics}, together with the derivation of
the effective Hamiltonian describing superconductivity of holes in semiconductors
induced by a completely general interband coupling from the proximity effect in
the conduction band. This result is then specialized in Sec.~\ref{sec:bulk} to the
case of the Kane-model description~\cite{kan57,win03,yu10} of typical
semiconductors. A comprehensive study of resulting changes to the electronic
valence-band structure in bulk systems and various types of nanostructures is
presented in Sec.~\ref{sec:nano}. We provide a summary and conclusions of
our work in Sec.~\ref{sec:concl}. Some mathematical details are given in the
Appendix.

\section{Basic theory and general results\label{sec:basics}}

In order to derive the induced superconducting pair potential for the
valence-band holes, we consider as a starting point the multi-band BdG
Hamiltonian for a superconductor-semiconductor hybrid structure~\cite{fut11},
\begin{equation}\label{eq:BdG12}
{\mathcal H}_{\text{BdG}} = \left( \begin{array}{cccc} \ham_{\text{c}} -\mu
& \ham_{\text{c-v}} & \Delta \, \openone_{2\times 2} & 0_{2\times 4} \\
\ham_{\text{c-v}}^\dagger & \ham_{\text{v}} - \mu & 0_{4\times 2} &
0_{4 \times 4} \\ \Delta^\ast \, \openone_{2\times 2} & 0_{2\times 4} & \mu
- \ham_{\text{c}} & - \ham_{\text{c-v}} \\ 0_{4\times 2} & 0_{4\times 4} &
- \ham_{\text{c-v}}^\dagger & \mu - \ham_{\text{v}}
\end{array} \right) \,\, ,
\end{equation}
where $ \ham_{\text{c(v)}}$ is the effective-mass Hamiltonian of the conduction
(valence) band, $\ham_{\text{c-v}}$ describes the coupling between conduction
and valence-band states, and $\Delta$ is the pair potential induced for
conduction-band states only via contact to a simple-metal superconducting
material. The matrix $\openone_{m \times m}$ is the identity matrix of dimension
$m\times m$, and $0_{m \times n}$ is the zero matrix of dimension $m \times n$.
The dimensionality for sub-blocks of ${\mathcal H}_{\text{BdG}}$ given in
Eq.~(\ref{eq:BdG12}) is determined by the fact that charge carriers from the
conduction (valence) band carry a spin-1/2 (spin-3/2) degree of freedom~\cite{yu10}.
We adopt the representation where eigenstates for spin projections on the
$z$ axis comprise the basis and use the order $\ket{\mathrm{c}, +1/2}$,
$\ket{\mathrm{c}, -1/2}$ ($\ket{\mathrm{v}, +3/2}$, $\ket{\mathrm{v}, 1/2}$,
$\ket{\mathrm{v}, -1/2}$, $\ket{\mathrm{v}, -3/2}$) for spinor amplitudes of
conduction-band (valence-band) states. The explicit form of the conduction and
valence-band Hamiltonians depends on the particular model under consideration
but, for typical semiconductor materials, the inter-band coupling is quite generally
of the form~\cite{yu10}
\begin{equation}\label{eq:KaneCoupl}
\ham_{\text{c-v}} = \begin{pmatrix} -\frac{1}{\sqrt{2}} P k_+ & \sqrt{\frac{2}{3}}
P k_z & \frac{1}{\sqrt{6}} P k_- & 0\\ 0 & -\frac{1}{\sqrt{6}} P k_+ &
\sqrt{\frac{2}{3}} P k_z & \frac{1}{\sqrt{2}} P k_- \end{pmatrix} \, .
\end{equation}
Here  $P$ is the materials-dependent Kane-model~\cite{kan57} matrix element,
and $k_\pm \equiv k_x \pm i\, k_y$ in terms of Cartesian coordinates of the
band-electron wave vector $\kk$.

We can treat the inter-band coupling in lowest-order perturbation theory without
needing to consider its explicit form, and thus derive a general effective Hamiltonian
for the valence bands, by employing the L\"owdin-partitioning method~\cite{win03}.
For completeness, and to motivate further approximations, we briefly sketch
details of the calculation here. The BdG Hamiltonian [Eq.~(\ref{eq:BdG12})] is
split into a part $\mathcal{H}_0$ that describes the individual conduction and
valence bands, and a part $\mathcal{H}_1$ that embodies the mixing between
conduction and valence bands. We then have $ {\mathcal H}_{\text{BdG}}=
\mathcal{H}_0+\mathcal{H}_1$, with 
\begin{align}\label{eq:pertHam1}
\mathcal{H}_1 = \left( \begin{array}{cccc} 0_{2\times 2}
& \ham_{\text{c-v}} & 0_{2\times 2} & 0_{2\times 4} \\
\ham_{\text{c-v}}^\dagger & 0_{4\times 4} & 0_{4\times 2} &
0_{4 \times 4} \\ 0_{2\times 2} & 0_{2\times 4} &  0_{2\times 2}
& - \ham_{\text{c-v}}\\ 0_{4\times 2} & 0_{4\times 4} & - \ham_{\text{c-v}}^\dagger
& 0_{4 \times 4} \end{array} \right) \,\, .
\end{align} 
The effective Hamiltonian for valence-band states that accounts for the presence
of inter-band coupling can be found by performing a unitary transformation to eliminate
the perturbation $\mathcal{H}_1$;
\begin{align*}
 &\tilde{{\mathcal H}}_{\text{BdG}}=\ee^{-S}~{\mathcal H}_{\text{BdG}}~ \ee^{S}\\
 &\approx\mathcal{H}_0+\mathcal{H}_1+[\mathcal{H}_0,S]+[\mathcal{H}_1,S]
 +\frac{1}{2}[[\mathcal{H}_0,S],S]+\mathcal{O}(\mathcal{H}_1^3 )~.
\end{align*}
This is generally a perturbative procedure that can be carried out to any desired
order~\cite{fol50,loe51,win03}. For our purposes, it will be sufficient to eliminate
$\mathcal{H}_1$ in first order, thus the generator $S$ needs to fulfill the condition
\begin{align}
\label{eq:S-condition}
\mathcal{H}_1+[\mathcal{H}_0,S]=0 \quad .
\end{align}
Using for $S$ the solution of Eq.~(\ref{eq:S-condition}) yields for the transformed
Hamiltonian
\begin{align}\label{eq:genDiagHam}
\tilde{{\mathcal H}}_{\text{BdG}}\approx\mathcal{H}_0
+\frac{1}{2}[\mathcal{H}_1,S] \quad . 
\end{align}

So far, we have not specified a particular form of the effective-mass Hamiltonians
$\ham_{\text{c}}$ and $\ham_{\text{v}}$ for the conduction and valence-band carriers.
To be consistent with the widely used Kane and Luttinger-model descriptions for typical
semiconductors~\cite{lut56,kan57,lip70,suz74,tre79,win03,yu10}, we will from now on
only keep terms upto quadratic order in components of $\kk$ in any Hamiltonian's matrix
elements. As a result, when solving Eq.~(\ref{eq:S-condition}) for the generator $S$ of
the unitary transformation, we can set all energy differences between conduction and
valence-band states equal to the fundamental band gap $E_{\text{g}}$, as the neglected
wave-vector dependences would ultimately lead to corrections that are of
higher-than-quadratic order in components of $\kk$.~\footnote{Fundamentally,
substituting $E_{\text{g}}$ for energy denominators in $S$ is justified because
$\mathcal{H}_1$ is linear in components of $\kk$. See Eqs.~(\ref{eq:KaneCoupl}) and
(\ref{eq:pertHam1}). By virtue of Eq.~(\ref{eq:S-condition}), $S$ must then be linear
in components of $\kk$ to leading order and, hence, the leading wave-vector dependence
of the commutator in Eq.~(\ref{eq:genDiagHam}) is quadratic. Therefore,
$\kk$-dependent terms from energy denominators in $S$ will only give rise to corrections
of higher-than-quadratic order in wave-vector components. As terms of this type are
neglected in commonly used expressions for $\mathcal{H}_0$, the form of $S$
given in Eq.~(\ref{eq:S-generator}) is adequate.}
Furthermore, in the spirit of the L\"owdin-partitioning calculation, we assume $|\mu|\ll
E_{\text{g}}$, $|\Delta| \ll E_{\text{g}}$, and also neglect corrections of order
$|\Delta|^2/E_{\text{g}}^2$. Within  this framework, we find  
\begin{align}\label{eq:S-generator}
S =\frac{1}{E_{\text{g}}} \left( \begin{array}{cccc}
 0_{2\times 2}
& -\ham_{\text{c-v}} & 0_{2\times 2} &\frac{\Delta}{E_{\text{g}}}\ham_{\text{c-v}} \\
\ham_{\text{c-v}}^\dagger & 0_{4\times 4} & \frac{\Delta}{E_{\text{g}}}\ham_{\text{c-v}}^\dagger 
& 0_{4 \times 4} \\ 
0_{2\times 2} & -\frac{\Delta^*}{E_{\text{g}}}\ham_{\text{c-v}} &  0_{2\times 2} &
- \ham_{\text{c-v}}\\ 
-\frac{\Delta^*}{E_{\text{g}}}\ham_{\text{c-v}}^\dagger & 0_{4\times 4} &
 \ham_{\text{c-v}}^\dagger & 0_{4 \times 4}
\end{array} \right) \,\, .
\end{align}
Using Eq.~(\ref{eq:S-generator}), the transformed Hamiltonian is found as
\begin{widetext}
\begin{align}
\label{transfH}
\tilde{{\mathcal H}}_{\text{BdG}}=\mathcal{H}_0+\frac{1}{E_{\text{g}}} \left(
\begin{array}{cccc}
 \ham_{\text{c-v}} \ham_{\text{c-v}}^\dagger 
&   0_{2\times 4} &\frac{\Delta}{E_{\text{g}}}\ham_{\text{c-v}}\ham_{\text{c-v}}^\dagger
&0_{2\times 4} \\
0_{4\times2} &- \ham_{\text{c-v}}^\dagger \ham_{\text{c-v}} &0_{4 \times 2}  &
\frac{\Delta}{E_{\text{g}}} \ham_{\text{c-v}}^\dagger \ham_{\text{c-v}} \\ 
 \frac{\Delta^*}{E_{\text{g}}}\ham_{\text{c-v}}\ham_{\text{c-v}}^\dagger&0_{2\times 4} &
 - \ham_{\text{c-v}} \ham_{\text{c-v}}^\dagger  & 0_{2\times 4}\\ 
0_{4\times2} &  \frac{\Delta^*}{E_{\text{g}}} \ham_{\text{c-v}}^\dagger \ham_{\text{c-v}}
& 0_{4 \times 2}  & \ham_{\text{c-v}}^\dagger \ham_{\text{c-v}} 
\end{array} \right) \,\, .
\end{align}
\end{widetext}
By construction there is no direct coupling between valence and conduction bands 
in the transformed Hamiltonian Eq.~(\ref{transfH}). However, a pair potential has
now been generated in the valence bands that is, to lowest order, quadratic in the
perturbation $\mathcal{H}_1$. Equation~(\ref{transfH}) forms the starting point for
our further analysis of the induced pair potentials for valence-band states in
various kinds of superconductor-semiconductor hybrid systems. 

\begin{figure*}[t]
\includegraphics[width=0.45 \textwidth]{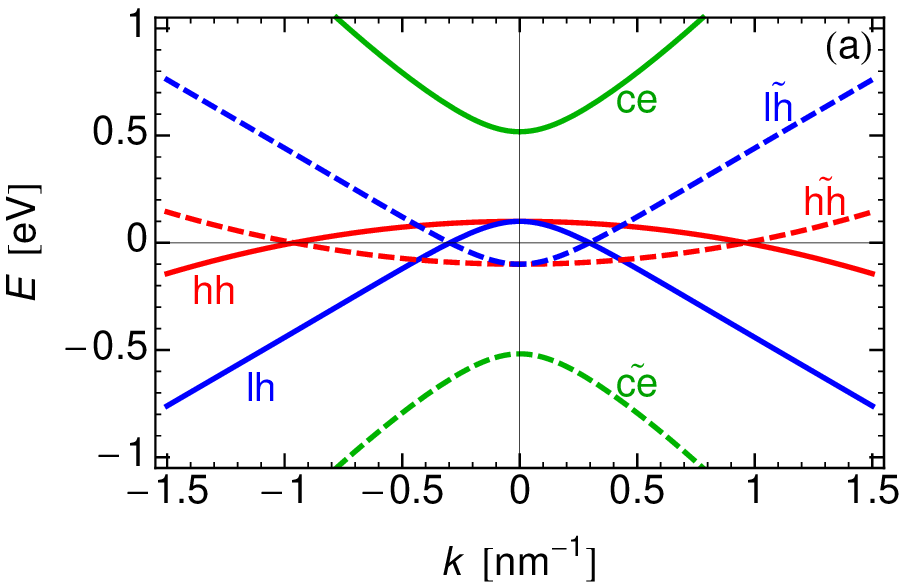}\hspace{1cm}
\includegraphics[width=0.45 \textwidth]{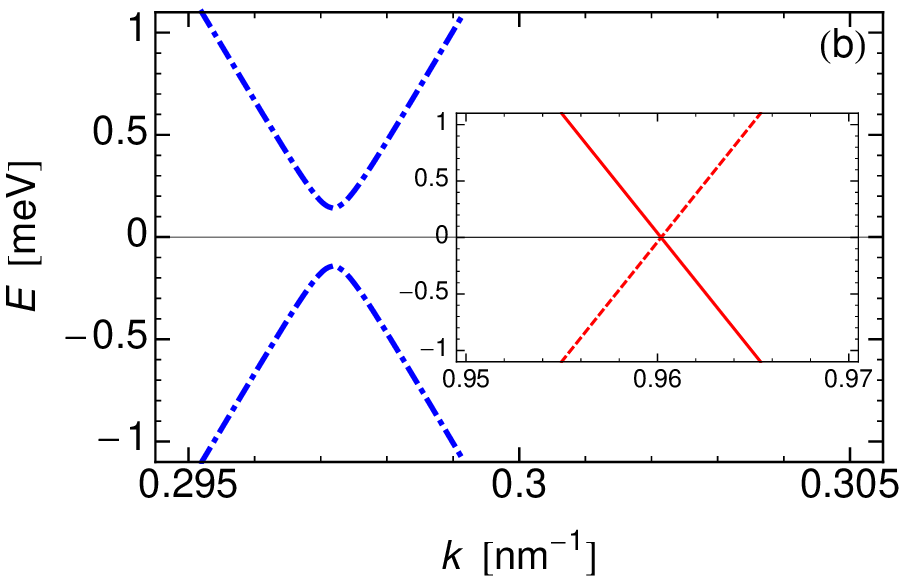}
\caption{\label{fig:Dispersions}Energy dispersions for Bogoliubov quasiparticles.
Panel (a) shows bands for conduction electrons (ce), heavy holes (hh), light holes
(lh), and their respective time-reversed partners (distinguished by a tilde) for
$\Delta=0$. In the calculation, band-structure parameters for InAs have been
used~\cite{win03} ($E_{\text{g}}=0.418$~eV, $P=9.197$~eV\,\AA,
$m^*=0.0229~m_0$, $\gamma_1=20.4$, $\gamma_2=8.3$, $\gamma_3=9.1$),
and a chemical potential $\mu=-0.1$~eV has been assumed. Panel (b) shows a
zoom-in near the chemical potential for the (experimentally realistic~\cite{chr97})
case with $\Delta=1$~meV. Note that there is no coupling between heavy holes
and their time-reversed partner excitations, whereas the familiar superconducting
gap appears in the excitation spectrum of the light holes.}
\end{figure*}

\section{Proximity-induced effective pair potential for the bulk valence band\label{sec:bulk}}

Results obtained in the previous Section enable the explicit derivation of an
effective BdG Hamiltonian for the upper-most valence band. Neglecting
${\mathcal O} (|\Delta|^2/E_{\text{g}}^2)$ corrections to effective-mass parameters
and, in the spirit of the usual $\kk$dot$\vek{p}$ approach~\cite{yu10}, keeping only
terms upto quadratic order in $\kk$, we find
\begin{equation}\label{eq:vbBdG}
{\mathcal H}_{\text{BdG}}^{(\text{eff})} = \left( \begin{array}{cc}
\ham_{\text{v}}^{(\text{eff})} - \mu & {\mathsf\Delta}_{\text{v}}^{(\text{eff})} \\
\left({\mathsf\Delta}_{\text{v}}^{(\text{eff})}\right)^\dagger & \mu -
\ham_{\text{v}}^{(\text{eff})} \end{array} \right) \quad ,
\end{equation}
where $\ham_{\text{v}}^{(\text{eff})}$ is the $4\times 4$ Luttinger-model
Hamiltonian~\cite{lut56} describing the bulk valence band in the uniform
semiconductor material, and
\begin{subequations}\label{eq:Gapmatrix}
\begin{eqnarray}
{\mathsf\Delta}_{\text{v}}^{(\text{eff})} &=&  \frac{\Delta}{E_{\text{g}}^2} \,\,
\ham_{\text{c-v}}^\dagger\cdot\ham_{\text{c-v}} \quad , \\ \label{eq:vbPairPot}
&\equiv& \Delta \, \frac{\bar\gamma_1\hbar^2}{2 m_0 E_{\text{g}}} \left[
\frac{9}{4}\, \kk^2 \, \openone_{4\times 4} - \left( \kk\cdot \vek{J} \right)^2
\right] \quad.
\end{eqnarray}
\end{subequations}
Here $\vek{J}=(J_x, J_y, J_z)$ denotes the vector of spin-3/2 matrices~\cite{win03}
that satisfy the usual angular-momentum commutation relations, and
$\bar\gamma_1 \equiv 2 m_0 P^2/(3\hbar^2 E_{\text{g}})$ is an effective-mass
parameter familiar from the Kane model~\cite{kan57}. Note that the structure
of the $\kk$-dependent terms in Eq.~(\ref{eq:vbPairPot}) coincides with that
for a Luttinger-model Hamiltonian where $\gamma_2 = \gamma_3 = \bar
\gamma_1/2$.

Two features exhibited by the pair potential ${\mathsf\Delta}_{\text{v}}^{(\text{eff})}$
in the valence band are remarkable and have not been discussed before:
(i)~its prominent (leading-order) $\kk$ dependence, and (ii)~its matrix structure
in the valence-band (spin-3/2-projection) subspace. As discussed in greater
detail below, property (i) results in characteristic features for the proximity
effect in quantum-confined structures. Property (ii) in conjunction with (i)
implies that the electronic excitations in a valence band with proximity-induced
superconducting correlations will be mixtures of heavy-hole (HH) and light-hole
(LH) amplitudes~\footnote{We use the usual nomenclature where hole states
corresponding to spin projections $\pm 3/2$ and $\pm 1/2$ are called heavy
holes and light holes, respectively.}. In this context, it is instructive to
relate property (ii) to the most general possible form of a pair potential
between spin-3/2 particles~\cite{sig91}, which is given by 
\begin{align}\label{eq:s-d-wave}
{\mathsf\Delta}_{3/2}^{(\text{eff})} =  \left( \begin{array}{cccc}
 \Delta_{\frac{3}{2},-\frac{3}{2}}
&  \Delta_{\frac{3}{2},-\frac{1}{2}} &  \Delta_{\frac{3}{2},\frac{1}{2}} &  \Delta_{\frac{3}{2},\frac{3}{2}} \\
  \Delta_{\frac{1}{2},-\frac{3}{2}}
&  \Delta_{\frac{1}{2},-\frac{1}{2}} &  \Delta_{\frac{1}{2},\frac{1}{2}} &  \Delta_{\frac{1}{2},\frac{3}{2}} \\
 \Delta_{-\frac{1}{2},-\frac{3}{2}}
&  \Delta_{-\frac{1}{2},-\frac{1}{2}} &  \Delta_{-\frac{1}{2},\frac{1}{2}} &  \Delta_{-\frac{1}{2},\frac{3}{2}} \\
 \Delta_{-\frac{3}{2},-\frac{3}{2}}
&  \Delta_{-\frac{3}{2},-\frac{1}{2}} &  \Delta_{-\frac{3}{2},\frac{1}{2}} &  \Delta_{-\frac{3}{2},\frac{3}{2}} \\
\end{array} \right) \,\, .
\end{align}
As time-reversal symmetry is intact in our system of interest, the relation
$\Delta_{i,j}=\Delta_{j,i}$ holds. Furthermore, parity is a good symmetry also,
restricting pairing amplitudes further to be of singlet type~\cite{sig91} and,
thus, an even function of $\kk$. The most general form of the pair-potential
matrix elements for a spin-3/2 degree of freedom therefore mirrors that
of the $\kk$dot$\vek{p}$ Hamiltonian~\cite{lip70,suz74,tre79} elements. In
terms of the familiar notation of \textit{s}, \textit{p}, \textit{d}, \textit{f}, $\dots$
contributions to superconducting-pair amplitudes, we have
\begin{align*}
 \Delta_{\frac{3}{2},-\frac{3}{2}}: \,\, s-d_{z^2}~, \qquad & \Delta_{\pm\frac{3}{2},
 \pm \frac{1}{2}}: \,\, d_{x^2-y^2}\mp id_{xy}~, \\
 \Delta_{\frac{1}{2},-\frac{1}{2}}: \,\, s+d_{z^2}  ~, \qquad & \Delta_{\pm\frac{3}{2},
 \mp \frac{1}{2}}: \,\, d_{xz}\pm id_{yz}~.
\end{align*}
The fact that ${\mathsf\Delta}_{\text{v}}^{(\text{eff})}$ from Eq.~(\ref{eq:Gapmatrix})
exhibits spherical symmetry in $\kk$ space is a consequence of the particular (Kane)
model utilized in our approach. It can be expected that, in the most general case, the
proximity-induced pair potential in the valence band will only be constrained by the
cubic lattice symmetry and, hence, can be any linear combination of symmetry-allowed
contributions shown in Table~II of Ref.~\onlinecite{sig91}. 

\section{Dimensional dependence of the proximity effect in the valence band\label{sec:nano}}

Based on the effective BdG Hamiltonian (\ref{eq:vbBdG}) for the uppermost valence
band, we now investigate how the electronic properties of holes are changed due to
the proximity effect. In particular, we focus on how the effective pair potential and the
induced superconducting gap depend on the dimensionality of the hole system. 

\subsection{Three-dimensional (bulk) systems}

In typically applied $\kk$dot$\vek{p}$ models~\cite{win03}, the eigenstates of the
effective valence-band Hamiltonian $\ham_{\text{v}}^{(\text{eff})}$ are the
spin-3/2-projection eigenstates (HHs and LHs) when the spin-quantization ($z$)
axis is taken to be parallel to $\kk$. It can be straightforwardly seen that the effective
pair potential (\ref{eq:vbPairPot}) is also diagonal in this representation, hence the
bulk-valence-band BdG Hamiltonian is block-diagonal in the HH and LH degrees of
freedom,
\begin{equation}
{\mathcal H}_{\text{BdG}}^{(\text{eff},\text{3D})} = \bigotimes_{\lambda=
\pm 3/2, \pm 1/2} \begin{pmatrix} E^{(\lambda)}_\kk - \mu & \Delta^{(\lambda)}
\\\left( \Delta^{(\lambda)}\right)^* & \mu - E^{(\lambda)}_\kk \end{pmatrix} \quad .
\end{equation}
Here $E^{(\lambda)}_\kk$ for $\lambda=\pm 3/2$ ($\lambda=\pm 1/2$) is the
bulk HH (LH) dispersion, and
\begin{equation}\label{bulk-delta}
\Delta^{(\pm 3/2)}=0 \,\,\, , \,\,\, \Delta^{(\pm 1/2)} = \Delta \,\, \frac{\bar\gamma_1
\hbar^2 k^2}{m_0 E_{\text{g}}} \quad .
\end{equation}
We thus find that the induced superconducting gap/pair potential vanishes for
the HH bands, whereas a finite gap exists for the LH bands. The latter's $\kk$
dependence implies a scaling with hole density $\varrho_{\text{3D}}$ as
$\varrho_{\text{3D}}^{2/3}$. The finding of vanishing pair correlations for HHs is
consistent with the absence of Andreev reflections found in Ref.~\onlinecite{fut11}
for HHs with perpendicular incidence on the interface with a superconductor. 

The characteristics of the proximity effect induced in the valence band via
coupling to the conduction band are illustrated in Fig.~\ref{fig:Dispersions}.
We plot dispersions of relevant energy bands obtained from the $12\times 12$
BdG Hamiltonian in Eq.~(\ref{eq:BdG12}), with band-structure parameters
applicable for InAs~\cite{vur01,win03}. To set the scene, Fig.~\ref{fig:Dispersions}(a)
shows the result for $\Delta=0$, where conduction-electron states (ce),
light-hole states (lh) and heavy-hole states (hh)  are decoupled from their
corresponding time-reversed states (denoted by $\widetilde{\text{ce}}$,
$\widetilde{\text{lh}}$ and $\widetilde{\text{hh}}$, respectively). 
Figure.~\ref{fig:Dispersions}(b) zooms in on the valence-band dispersions
in the vicinity of the Fermi level when $\Delta=1$~meV. In agreement with
(\ref{bulk-delta}), an energy gap is found for the LH-like states and, in the
vicinity of the Fermi level, the LH-like quasiparticle states are linear combinations
of normal excitations (${\text{ce}}$ and ${\text{lh}}$)  and time-reversed partners
($\widetilde{\text{ce}}$ and $\widetilde{\text{lh}}$). In contrast, no gap appears for
the HH states. Thus, in summary, the proximity effect in bulk-hole systems can be
understood in terms of the familiar decoupled HH and LH excitations, with HHs
being unaffected and LHs experiencing superconductivity with an $s$-wave type
pairing that, to leading order, depends quadratically on wave vector
($\Delta_{\frac{1}{2},-\frac{1}{2}}\propto |\kk|^2$).

\subsection{Quasi-twodimensional (quantum-well) systems}

A quantum-well confinement of electrons in the semiconductor structure can be
treated by replacing $\ham_{\text{v}}^{(\text{eff})}\to \ham_{\text{v}}^{(\text{eff})}
+ V(z)$ in Eq.~(\ref{eq:vbBdG}). Subband-$\kk$dot$\vek{p}$ theory~\cite{bro85,yan85}
could then be applied to find the energy dispersions and the corresponding eigenstates.
To get a qualitative insight into the proximity effect for quasi-twodimensional
(quasi-2D) hole systems, we treat the confining potential in an approximate
manner~\cite{ker10} by setting $k_z\to 0$ and $k_z^2 \to n^2\pi^2/d^2$ in the
bulk-hole BdG Hamiltonian, Eq.~(\ref{eq:vbBdG}). Here $d$ denotes the effective
quantum-well width, and $n=1, 2, \dots$ labels orbital bound states associated
with the quantum-well potential $V(z)$. Such a procedure renders the valence-band
BdG Hamiltonian block-diagonal for each $n$, with $2\times 2$ blocks in the
subspaces $\{3/2,-1/2\}$ and $\{- 3/2, 1/2\}$ labelled by the index $\sigma=\pm 1$:
\begin{equation}\label{eq:BdG2D}
{\mathcal H}_{\text{BdG}}^{(\text{eff},\text{2D})} = \bigotimes_{n=1,2,\dots
\atop \sigma=\pm 1} \begin{pmatrix} \ham_{\sigma, n}^{(\text{2D})} - \mu\,
\openone_{2\times 2} & \Delta_{\sigma, n}^{(\text{2D})} \\ \left(\Delta_{\sigma,
n}^{(\text{2D})} \right)^\dagger& \mu \, \openone_{2\times 2} - \ham_{\sigma, n}^{(\text{2D})}
\end{pmatrix} .
\end{equation}
Defining $\vek{k}_\perp = (k_x, k_y)$ and adopting the spherical approximation
for the Luttinger model~\cite{suz74,tre79}, the sub-matrices are given by
\begin{subequations}
\label{submats}
\begin{eqnarray}
\ham_{\sigma, n}^{(\text{2D})} &=&\begin{pmatrix}  E_n^{(\text{h})} - \frac{\hbar^2
k_\perp^2}{2 m_\perp^{(\text{h})}} & \frac{\sqrt{3}\gamma_3 \hbar^2 k_{-\sigma}^2}{2 m_0} \\
\frac{\sqrt{3}\gamma_3\hbar^2 k_\sigma^2}{2 m_0} & E_n^{(\text{l})} - \frac{\hbar^2
k_\perp^2}{2 m_\perp^{(\text{l})}} \end{pmatrix} \,\, , \\[0.3cm]
\Delta_{\sigma, n}^{(\text{2D})} &=& \Delta \, \frac{\bar\gamma_1 \hbar^2}{2 m_0
E_{\text{g}}} \begin{pmatrix} \frac{3 k_\perp^2}{2}  & - \frac{\sqrt{3}k_{-\sigma}^2}{2}
\\[0.2cm] - \frac{\sqrt{3}k_\sigma^2}{2} & \frac{2\pi^2 n^2}{d^2} + \frac{k_\perp^2}{2}\,
\end{pmatrix} \,\, ,
\end{eqnarray}
\end{subequations}
where $E_n^{(\text{h/l})}$ are quasi-2D subband energies for HH/LH states at
$\kk_\perp=0$, and $m_\perp^{(\text{h/l})}$ their respective effective masses for the
in-plane motion. For a $(001)$ quantum well, these effective masses are
$m_\perp^{(\text{h/l})} = m_0/(\gamma_1 \pm \gamma_2)$ in terms of the
standard~\cite{vur01} Luttinger parameters.

For the sake of completeness, we have given the expressions of the sub-matrices
Eqs.~(\ref{submats}) for an arbitrary value of the orbital bound-state index $n$.
However, it should be noted that the results are reliable only for the lowest subband,
i.e., $n=1$. The Hamiltonian~(\ref{eq:BdG2D}) together with Eqs.~(\ref{submats}) is
the central result of this section; it provides a simple model to describe
proximity-induced superconductivity in quasi-2D hole systems in the low-density
regime (such that only one quantum-well subband is occupied).  

\begin{figure}[b]
\includegraphics[width=0.9\columnwidth]{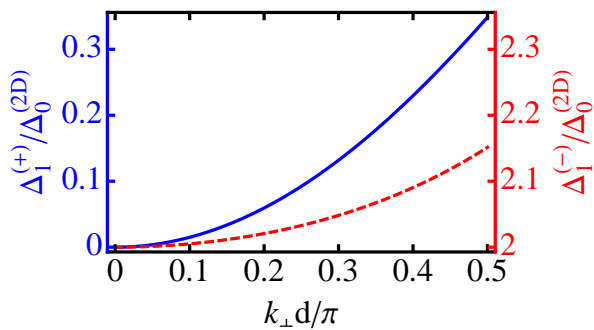}
\caption{\label{fig:2Dgap}Effective superconducting pair potentials of quasi-2D
holes. The result for HH-like (LH-like) excitations in the $n=1$ quantum-well
bound state, i.e., the lowest (first excited) subband, is shown as the
blue solid (red dashed) curve. See also Eq.~(\ref{eq:2Dord}). The different scales
for $\Delta^{(+)}_1$ (blue/left ordinate) and $\Delta^{(-)}_1$ (red/right ordinate)
emphasize the different band widths in the $\kk$-dependence for pair potentials
of HH-like and LH-like excitations, as well as the absence/existence of a constant
contribution. Band-structure parameters have been absorbed into the definition of
$\Delta^{(2\text{D})}_0\equiv\pi^2\bar\gamma_1\hbar^2\Delta/(2m_0 d^2
E_{\text{g}})$.}
\end{figure}

To make contact with the usual intuition about superconducting correlations,
we perform a unitary transformation of ${\mathcal H}_{\text{BdG}}^{(\text{eff},
\text{2D})}$ that diagonalizes the pair-potential matrices
$\Delta_{\sigma, n}^{(\text{2D})}$. Straightforward calculation yields
\begin{subequations}
\begin{equation}
\tilde \Delta_{\sigma, n}^{(\text{2D})} \equiv {\mathcal U}\, \Delta_{\sigma, n}^{(\text{2D})}
\, {\mathcal U}^\dagger = \begin{pmatrix} \Delta_n^{(+)} & 0 \\ 0 & \Delta_n^{(-)}
\end{pmatrix} \,\, ,
\end{equation}
where (using $\kappa := k_\perp d/\pi$)
\begin{equation}\label{eq:2Dord}
\Delta_n^{(\pm)} = \Delta \, \frac{\pi^2 \bar\gamma_1 \hbar^2}{2 m_0 d^2 E_{\text{g}}}
\left( n^2 + \kappa^2 \mp \sqrt{n^4 - \kappa^2 + \kappa^4} \right)\, .
\end{equation}
\end{subequations}
Note the limiting behavior for $\kappa\ll 1$:
\begin{subequations}
\begin{eqnarray}
\Delta_n^{(+)} &\to& \Delta\, \frac{3 \bar\gamma_1 \hbar^2 k_\perp^2}{4 m_0
E_{\text{g}}}~,\\[2mm]
\Delta_n^{(-)} &\to& \Delta \, \frac{\pi^2 \bar\gamma_1\hbar^2 n^2}{m_0 d^2
E_{\text{g}}}\left[1+\left(\frac{k_\perp d}{2\pi}\right)^2\right]~ .
\end{eqnarray}\label{eq:2Dgap}
\end{subequations}
In Fig.~\ref{fig:2Dgap}, we plot the induced pair potentials from Eq.~(\ref{eq:2Dord})
as a function of $k_\perp d/\pi$. As the same transformation that diagonalizes
$\Delta_{\sigma, n}^{(\text{2D})}$ also renders $\ham_{\sigma, n}^{(\text{2D})}$ to
be diagonal within the spherical approximation, we find that the effective pair potential
for quasiparticles from the HH-like quasi-2D hole subbands is $k_\perp$-dependent (i.e,
does not vanish as was the case for HH states in the bulk) and can thus strongly
increase with variation of the quasi-2D hole density. Also unlike their bulk counterparts,
the LH-like quasi-2D hole states are subject to a pair potential that is, to leading order,
a constant. In particular, states at the Fermi energy in the lowest subband, which is
HH-like, will have a density-dependent superconducting gap that scales linearly with
the quasi-2D hole density. Except at $\kk_\perp=0$, the quasiparticle states arising from
superconducting correlations are mixtures of HH and LH states. State-dependent
physical quantities such as response functions will be affected by this HH-LH mixing,
reflecting the spinor structure of confined hole states~\cite{ker10,ker13,ker13a,ker13b}.

\subsection{Quasi-onedimensional (nanowire) systems}

We consider hole nanowires defined by a transverse hard-wall confinement for two
distinct sample geometries: 1)~a system with rectangular cross-section, which can
serve as a model for quantum wires obtained by electrostatic confinement of a 2D
hole gas~\cite{sri13}, and 2)~a cylindrical nanowire that can be fabricated, e.g., by
the VLS growth method~\cite{lu06}. 

\subsubsection{Quantum wire with rectangular cross-section}

We first consider the holes to be confined in a rectangular quasi-1D system by a
hard-wall confinement with potential $V(x,y)=V(x)V(y)$, where the quantum-well
widths in $x$ and $y$ directions are $d$ and $w$, respectively. We use again
an approximation, setting $k_{x}, k_{y}\to 0$ and $k_{x}^2\to n^2 \pi^2/d^2$ and
$k_{y}^2\to n'^2 \pi^2/w^2$. Similarly to Eq.~(\ref{eq:BdG2D}), the corresponding
BdG Hamiltonian is then block-diagonal, with associated $2\times2$ sub-matrices
\begin{subequations}
\begin{eqnarray}
\ham_{\sigma, n n'}^{(\text{1D})} &=& \nonumber \\[0.2cm] && \hspace{-1.5cm}
\begin{pmatrix}
- \frac{\pi^2\hbar^2 (n^2+r^2n'^2)}{2 m_{\perp}^{(\text{h})}d^2} -
\frac{\hbar^2 k_z^2}{2 m_\|^{(\text{h})}} & \frac{\sqrt{3}\gamma_3\pi^2
\hbar^2 (n^2 - r^2 n'^2)}{2 m_0d^2} \\ \frac{\sqrt{3}\gamma_3\pi^2\hbar^2
(n^2 - r^2 n'^2)}{2 m_0 d^2} & - \frac{\pi^2\hbar^2 (n^2+r^2n'^2)}{2
m_{\perp}^{(\text{l})}d^2} - \frac{\hbar^2 k_z^2}{2 m_\|^{(\text{l})}}
\end{pmatrix} , \quad \nonumber \\ \\[0.3cm]
\label{Delta-1drect}
\Delta_{\sigma, n n'}^{(\text{1D})} &=& \Delta \, \frac{\pi^2 \bar\gamma_1 \hbar^2}{2
m_0 d^2 E_{\text{g}}} \nonumber \\ && \hspace{-0.3cm} \times \begin{pmatrix}
\frac{3}{2}(n^2 + r^2 n'^2) & - \frac{\sqrt{3}}{2} (n^2 - r^2 n'^2) \\[0.2cm] - \frac{\sqrt{3}}{2}
(n^2 - r^2 n'^2) & \frac{1}{2}(1+r^2)+2 \kappa_z^2\, \end{pmatrix} , 
\end{eqnarray}
\end{subequations}
where $\kappa_z \equiv k_z d/\pi$, $r\equiv d/w$, and $m_\|^{(\text{h/l})}$ are
the effective masses for HH/LH motion parallel to the spin-3/2 quantization axis.
[For example, $m_\|^{(\text{h/l})} = m_0/(\gamma_1 \mp 2 \gamma_2)$ when the
quantization axis is parallel to the $(001)$ crystallographic direction.] The
index $\sigma=\pm$ distinguishes the subspaces $\{3/2,-1/2\}$ and $\{- 3/2, 1/2\}$,
respectively. From now on, we focus on the lowest quasi-1D hole subband, which
has $n=n'=1$. Diagonalizing Eq.~(\ref{Delta-1drect}) yields the induced
superconducting pair potential for this system,
\begin{eqnarray}
\Delta^{(\pm)} &=& \Delta \, \frac{\pi^2 \bar\gamma_1 \hbar^2}{2 m_0 d^2 E_{\text{g}}}
\left( 1+r^2+ \kappa_z^2 \right. \nonumber \\ && \hspace{0.5cm} \left.
\mp \sqrt{1-r^2+r^4- \kappa_z^2(1+r^2)+ \kappa_z^4 } \right)\, .
\end{eqnarray}
We recover the 2D result of Eq.~(\ref{eq:2Dord}) for $r\to 0$, with $k^2_\perp$
replaced by $k^2_z$. For the symmetric case $r=1$, we obtain
\begin{subequations}
\begin{eqnarray}
\left. \Delta^{(+)}\right|_{r=1} &=&\Delta\, \frac{\pi^2 \bar\gamma_1 \hbar^2}{2 m_0
d^2 E_{\text{g}}} \left[1+2 \left( \frac{k_z \pi}{d}\right)^2\right] , \\[2mm]
\left.\Delta^{(-)}\right|_{r=1} &=& \Delta\, \frac{3\pi^2 \bar\gamma_1 \hbar^2}{2 m_0
d^2 E_{\text{g}}} \,\, .
\end{eqnarray}
\end{subequations}
Again we find a nontrivial dependence of the induced superconducting pair potential
with respect to the hole density (quadratically dependent on the quantum wire hole
density) for the lowest subband, which is LH-like (for $r=1$) due to the phenomenon
of mass inversion, i.e., since $m_\perp^{(\text{h})} < m_\perp^{(\text{l})}$.

\subsubsection{Quantum wire with circular cross-section}

Next we consider a quasi-onedimensional (quasi-1D) system with cylindrical geometry.
In this case, it is convenient to use cylindrical coordinates $(r,\varphi,z)$. A hard-wall confinement
is  imposed by a potential $V(r)$ defined by $V(r)=0$ for $r<R$ and $V(r)=\infty$ for $r>R$,
where $R$ is the radius of the cylindrical wire. Due to the cylindrical symmetry of the system,
the Hamiltonian commutes with the projection of total angular momentum parallel to the
nanowire (i.e., the $z$) axis. Therefore the eigenstates of the Hamiltonian can be
classified~\cite{ser90} by the eigenvalue $\nu$ of $F_z = -i\partial_\varphi + J_z$. Using
the appropriate representation, we find the eigenfunctions at the subband edge by solving
the purely transverse bound-state problem ($k_z=0$). For $k_z=0$ the subspaces
$\{3/2,-1/2\}$ and $\{- 3/2, 1/2\}$ labelled by the index $\sigma=\pm 1$ are decoupled,
while for finite $k_z$ this is no longer the case.  We denote the subband-edge eigenfunctions
by  $\Psi^{(0)}_{(\sigma,\nu,n)}$, where the index $n$ labels the different radial quasi-1D
states associated to given values of $\sigma$ and $\nu$. The ground state is doubly degenerate
and, for wires made from InAs,~\footnote{The character of subband-edge states in cylindrical
quantum wires is strongly materials-dependent. See Ref.~\onlinecite{zue08} for a detailed
discussion. For example, in a GaAs nanowire, the quantum numbers for states at the
lowest (first excited) subband's edges are $\sigma=\pm 1,\nu=\mp 1/2, n=1$ ($\sigma=\pm 1,
\nu=\pm 1/2, n=1$), i.e., are those corresponding to the \emph{first excited (lowest)\/} subband's
edge in an InAs wire.} the two states have quantum numbers $\sigma=\pm 1,
\nu=\pm 1/2, n=1$, hence the corresponding wave functions for $k_z=0$ are
$\Psi^{(0)}_{(1,\frac{1}{2},1)}$ and  $\Psi^{(0)}_{(-1,-\frac{1}{2},1)}$. The
first excited subband is also doubly degenerate ($\sigma=\pm 1,\nu=\mp 1/2, n=1$),
and the wave functions are $\Psi^{(0)}_{(-1,\frac{1}{2},1)}$ and $\Psi^{(0)}_{(1,-\frac{1}{2},1)}$.
More details about the calculation of the cylindrical-hole-nanowire subbands are given in
the Appendix.

\begin{figure}[b]
\includegraphics[width=0.9 \columnwidth]{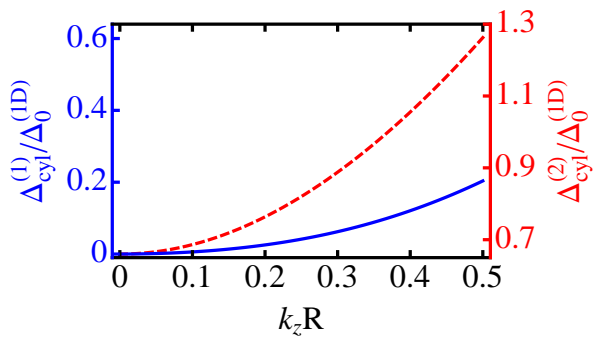}
\caption{\label{fig:gap1Dcyl}Effective superconducting pair potentials of quasi-1D hole
states in a cylindrical nanowire. See Eq.~(\ref{eq:Gapcylwire}). In the limit of small
$k_z R$, the blue solid (red dashed) curve is associated with eigenstates from the lowest
(first excited) nanowire subband. In general, the states that are directly coupled by these
pair potentials are superpositions of eigenstates at the same $k_z$ from the lowest two
subbands. We assumed band-structure parameters for InAs and defined
$\Delta^{(1\text{D})}_0\equiv\bar\gamma_1\hbar^2 \Delta/(2m_0 E_{\text{g}}R^2)$.}
\end{figure}

We project the nanowire Hamiltonian and pair potential onto the subspace of the two
lowest subband-edge states~\cite{ser90,cso09,klo11}. Since $\nu$ is a conserved
quantum number, the resulting Hamiltonian is block diagonal in the two subspaces
spanned by $\{\Psi^{(0)}_{(1,\frac{1}{2},1)}, \Psi^{(0)}_{(-1,\frac{1}{2},1)}\}$
and $\{\Psi^{(0)}_{(-1,-\frac{1}{2},1)}, \Psi^{(0)}_{(1,-\frac{1}{2},1)}\}$. In each
of the subspaces, the effective BdG Hamiltonian for the cylindrical quantum wire reads
\begin{subequations}
\begin{eqnarray}\label{eq:cylw1}
\ham_{\text{cyl}}^{(\text{1D})}&=& -\frac{\hbar^2\gamma_1}{2m_0 R^2}
\begin{pmatrix}  \tilde{E}^{(\text{g})} + \frac{
(k_z R)^2}{ \tilde{m}^{(\text{g})}} &i C k_z R\\ 
-i C k_z R & \tilde{E}^{(\text{e})} +\frac{
(k_z R)^2}{ \tilde{m}^{(\text{e})}}
\end{pmatrix},  \nonumber \\ \\[0.3cm]
\Delta_{\text{cyl}}^{(\text{1D})} &=& \frac{3\Delta\hbar^2\bar\gamma_1}{2m_0
R^2 E_{\text{g}}}
\begin{pmatrix}   \frac{
(k_z R)^2}{2} &i D k_z R\\ 
-i D k_z R&  \tilde{\Delta} + \frac{
(k_z R)^2}{\tilde{m}^{(\Delta)}}
\end{pmatrix}
  \, . \label{eq:cylw2}
\end{eqnarray}
\end{subequations}
For an InAs nanowire, the parameters entering Eqs.~(\ref{eq:cylw1}) and
(\ref{eq:cylw2}) are $\tilde{E}^{(\text{g})}=1.47$, $\tilde{E}^{(\text{e})}=2.23$,
$\tilde{m}^{(\text{g})}=0.69$, $\tilde{m}^{(\text{e})}=0.61$, $C=0.69$, $\tilde{\Delta}
=0.22$, $\tilde{m}^{(\Delta)}=1.74$, and $D=0.26$. Diagonalization of
Eq.~(\ref{eq:cylw2}) yields the induced superconducting pair potential of
hole-nanowire states to read
\begin{widetext}
\begin{equation}\label{eq:Gapcylwire}
\Delta_{\text{cyl}}^{(1,2)}=\frac{3\bar\gamma_1\hbar^2\Delta}{8m_0 R^2 E_{\text{g}}}
\left\{2\tilde\Delta+(k_zR)^2\left(1+\frac{2}{\tilde m^{(\Delta)}}\right) \mp
\sqrt{4\tilde\Delta^2+4(k_zR)^2\tilde\Delta\left(\frac{2}{\tilde m^{(\Delta)}}-1+
\frac{4D^2}{\tilde\Delta}\right)+(k_zR)^4\left(1-\frac{2}{\tilde m^{(\Delta)}}\right)^2}
\right\} .
\end{equation}
\end{widetext}
In the limit $k_zR\ll 1$, we find
\begin{subequations}
\begin{eqnarray}
\Delta_{\text{cyl}}^{(1)} &\to& \Delta \, \frac{3 \bar\gamma_1 \hbar^2}{2 m_0
E_{\text{g}}} \left(\frac{1}{2}-\frac{D^2}{\tilde\Delta}\right) k_z^2~, \\[2mm]
\Delta_{\text{cyl}}^{(2)} &\to& \Delta\, \frac{3 \bar\gamma_1 \hbar^2}{2 m_0R^2
E_{\text{g}}}\left[\tilde\Delta+\left(\frac{1}{\tilde{m}^{(\Delta)}}+\frac{D^2}{\tilde\Delta}
\right) \left(k_z R\right)^2 \right]~. \nonumber \\ 
\end{eqnarray}\label{eq:gap1Dcyl}
\end{subequations}
The states that are directly coupled by the superconducting pair potential turn out to
be mixtures of the two lowest-lying nanowire-subband states. The wave-vector
dependence of proximity-induced gap parameters is displayed in Fig.~\ref{fig:gap1Dcyl}.
It mirrors qualitatively the behavior of quantum-well states [cf.\ Eqs.~(\ref{eq:2Dord})
and (\ref{eq:2Dgap})], which is a point of difference with the rectangular-wire case. In
particular, the pair potential for lowest-subband-dominated states in a cylindrical InAs wire
is finite only when $k_z\neq 0$ and, therefore, strongly density-tunable. We note that the
effective Hamiltonian in Eq.~(\ref{eq:cylw1}) and the effective gap matrix, Eq.~(\ref{eq:cylw2}),
cannot be  simultaneously diagonalized in general.

\section{Summary and Conclusions\label{sec:concl}}

We have developed a general theoretical description of the superconducting proximity
effect for charge carriers from a semiconductor's valence band arising from coupling to
a simple-metal superconductor. Our starting point is an \textit{s}-wave pair potential that
is induced for conduction electrons by means of the ordinary proximity
effect~\cite{deg64,vol95,fag05,kop11}. We show how the familiar inter-band coupling
yields an unusual proximity effect for holes, with induced pair potentials depending
strongly on the dimensionality of the \textit{p}-doped system. In particular, in a bulk
sample, only light-hole state are subject to a finite pair potential, which depends
quadratically on wave vector. A rich behavior is revealed for quantum-confined
holes, with intriguing parallels being exhibited by quasi-2D (quantum-well) systems
and cylindrically shaped quantum wires made from InAs. See Figures \ref{fig:2Dgap}
and \ref{fig:gap1Dcyl}. In both of these cases, the pair potential couples states that
are mixtures between heavy-hole and light-hole components. The pair potential
affecting the lowest-lying subband states (shown as the blue solid curve in both figures)
is proportional to the squared magnitude of the hole wave vector and, therefore, strongly
tunable by the carrier density. In marked contrast to the bulk case, the pair potential for
quantum-confined holes can also have a constant contribution as is exhibited by the
first excited subbands (see the red dashed curve in both figures).

The wave-vector dependence of induced pair potentials for holes enables direct
tuning of superconducting correlations and gap parameters in these systems. 
This previously unnoticed feature renders confined hole systems ideal laboratories
for investigating the transition to, and properties of, novel topological phases with
their associated unconventional quasiparticle excitations~\cite{ali12,lei12,bee13,sta13}.
However, the influence of the, so far unappreciated, fact that the pair-potential-coupled
states are generally mixtures of heavy-hole and light-hole contributions requires further
detailed study. Also, a deeper understanding of the effect of Rashba-type
spin-orbit coupling and magnetic-field-induced spin splittings in conjunction with the
unconventional properties of proximity-induced superconductivity in hole systems
needs to be developed.

\begin{acknowledgments}
AGM would like to acknowledge useful discussions with J.~K\"onig at the early stages
of this work.
\end{acknowledgments}

\appendix*

\section{Luttinger-model description of cylindrical hole nanowires defined by a
hard-wall potential\label{sec:cylwire}}

To take full advantage of the cylindrical symmetry of the nanowire, we adopt the
cylindrical coordinates $(r,\varphi,z)$. The wire is described by the Schr\"odinger
equation $\left[\ham_{\text{v}}^{(\text{eff})} + V(r)\right] \Psi(\rr) = E\, \Psi(\rr)$. 
Within the axial approximation for the Luttinger Hamiltonian $\ham_{\text{v}}^{(\text{eff})}$,
the \textit{Ansatz\/}
\begin{equation}
\Psi(\rr) = \ee^{i k_z z}\,\, \ee^{i \left( \nu - J_z\right)\varphi}\,\, \Phi_\nu^{(k_z)}(r)
\end{equation}
transforms the original 3D Schr\"odinger problem for the nanowire into a 1D radial
equation $\left[\bar\ham_\nu(k_z) + V(r)\right] \Phi_\nu^{(k_z)}(r) = E_\nu(k_z)\,
\Phi_\nu^{(k_z)}(r)$. Here $\nu$ is the quantum number associated with the
total angular momentum component parallel to the wire axis, which is given
by $F_z = -i\partial_\varphi + J_z$.

In the spirit of subband $\kk$dot$\vek{p}$ theory, we separate the effective radial-motion
Hamiltonian into two parts, $\bar\ham_\nu(k_z) = \bar\ham_\nu^{(0)} + \bar\ham_\nu^{(1)}
(k_z)$, where $\bar\ham_\nu^{(0)} \equiv \bar\ham_\nu(k_z=0)$. Straightforward calculation
yields
\begin{widetext}
\begin{subequations}
\begin{eqnarray}\label{eq:Hamcylwire1}
\bar\ham_\nu^{(0)} &=& -\frac{\hbar^2}{2 m_0} \left\{ \left[ \gamma_1 + \gamma_2\left(
J_z^2 - \frac{5}{4}\,\openone\right)\right]\kk_{\| \nu}^2 + \gamma_3\left( J_+^2\, L_{\nu-1}\,
L_\nu + J_-^2\, L_{\nu+2}^\dagger\, L_{\nu+1}^\dagger \right) \right\} , \\
\bar\ham_\nu^{(1)}(k_z) &=& -\frac{\hbar^2}{2 m_0} \left\{ \left[ \gamma_1 -2 \gamma_2
\left( J_z^2 - \frac{5}{4}\,\openone\right)\right] k_z^2 + 2\sqrt{2}\, \gamma_2 \, i \left( \{
J_+\, , \, J_z \}\, L_\nu - \{ J_-\, , \, J_z \}\, L_{\nu+1}^\dagger \right) k_z \right\} .
\end{eqnarray}\label{eq:HamWire}
\end{subequations}
\end{widetext}
Here we have used the abbreviations
\begin{subequations}
\begin{eqnarray}
\kk_{\|\nu} &=& - \partial_r^2 - \frac{1}{r}\,\partial_r + \frac{\left(\nu - J_z \right)^2}{r^2}
\quad , \\ L_\nu &=& \partial_r + \frac{\nu - J_z}{r} \,\, , \,\, 
L_\nu^\dagger = -\partial_r + \frac{\nu - 1 - J_z}{r} \, . \quad
\end{eqnarray}
\end{subequations}
In the following, we adopt the spherical approximation, i.e. $\gamma_2=\gamma_3
\equiv\gamma_{\text{s}}=(2\gamma_2+3\gamma_3)/5$.  We first consider the
purely transverse motion, i.e. $k_z=0$. 

Straightforward inspection shows that $\bar\ham_\nu^{(0)}$ is block-diagonal in the
subspaces spanned by intrinsic angular-momentum projections $\{\pm 3/2, \mp 1/2\}$,
labelled by $\sigma=\pm 1$.  Therefore we can solve the transverse problem independently
for each value of $\sigma$. For this the transverse Schr\"odinger equation reads
\begin{equation}
\left[ \bar\ham_\nu^{(0)} + V(r) \right] \Phi^{(0)}_{\sigma, \nu, n}(r) = E_{\sigma, \nu, n}(0)\,
\Phi^{(0)}_{\sigma, \nu, n}(r) \quad ,
\end{equation}
where the index $n=1, 2, \dots$ labels in  order of ascending energies the different quasi-1D
subbands for given values of $\sigma$ and  $\nu$. Without confinement, the eigenstates for
the Hamiltonian (\ref{eq:Hamcylwire1}) are found to be
\begin{subequations}\label{eq:cylSymm2D}
\begin{align}
\phi_{1,\nu,\pm}(r)=
\left(
a_\pm\, J_{\nu-\frac{3}{2}}(k_\pm r),
0,
c_\pm\, J_{\nu+\frac{1}{2}}(k_\pm r),
0\right)^{T},\\
\phi_{-1,\nu,\pm}(r)=
\left(
0,
b_\pm\, J_{\nu-\frac{1}{2}}(k_\pm r),
0,
d_\pm\, J_{\nu+\frac{3}{2}}(k_\pm r)\right)^{T},
\end{align}
\end{subequations}
where $k_\pm=\sqrt{2m_0/(\hbar^2\gamma_1)}\sqrt{m_\pm |E|}$ with $E$ being
the energy, $m_\pm=1/(1\mp2\bar\gamma)$, $\bar\gamma=\gamma_{\text{s}}/
\gamma_1$, and coefficients $a_\pm=\{-1,\sqrt{3}\}$, $b_\pm=\{-\sqrt{3},1\}$, $c_\pm=
\{\sqrt{3},1\}$ and $d_\pm=\{1,\sqrt{3}\}$. In Eqs.~(\ref{eq:cylSymm2D}), Bessel
functions of the first kind (denoted by $J_M$) have been introduced. In the following,
we assume a quantum wire to be fabricated from InAs with the appropriate values for
the band structure parameters, which implies $\bar\gamma=0.45$. (The
materials dependence of valence-band states confined to cylindrical nanowires has
been discussed in Ref.~\onlinecite{zue08}.) The quantum-wire bound states
are written as a superposition of 2D states with the same value of $\sigma$, that is 
\begin{align}\label{eq:Boundstates}
\Phi_{\sigma,\nu,n}^{(0)}(r)=\frac{1}{R}\left[a_{\sigma,\nu,n}\phi_{\sigma,\nu,+}(r)+
b_{\sigma,\nu,n} \phi_{\sigma,\nu,-}(r)\right], 
\end{align}
with $\sigma=1$ or $-1$. 
The eigenergies are obtained by imposing the proper boundary condition at $r=R$. 
This amounts to solving the secular equation
\begin{align}
J_{\nu-\frac{3}{2}}(k_+ R)J_{\nu+\frac{1}{2}}(k_- R)+3 J_{\nu+\frac{1}{2}}(k_+ R)
J_{\nu-\frac{3}{2}}(k_- R)=0
\end{align}
for $\sigma=1$ and 
\begin{align}
J_{\nu+\frac{3}{2}}(k_+ R)J_{\nu-\frac{1}{2}}(k_- R)+3 J_{\nu-\frac{1}{2}}(k_+ R)
J_{\nu+\frac{3}{2}}(k_- R)=0 \nonumber\\[2mm]
\end{align}
for $\sigma=-1$. We find that the lowest subbands are the ones where ($\sigma=1,
\nu=\frac{1}{2}, n=1$) and ($\sigma=-1,\nu=-\frac{1}{2},n=1$). The associated bound
state energy is $|\varepsilon^{(1)}| = 1.47\,\varepsilon_0$ [with $\varepsilon_0=
\hbar^2\gamma_1/(2m_0 R^2)$], and the coefficients are $a_{-1,-1/2,1}=-a_{1,1/2,1}
= 0.70$ and $b_{1,1/2,1}=b_{-1,-1/2,1}= 0$, where the normalization condition for the
wave function has been included. 

The first excited subbands are the ones where $(\sigma=1,\nu= -\frac{1}{2},n=1)$ and
$(\sigma=-1,\nu=\frac{1}{2},n=1)$. The corresponding bound-state energy is
$|\varepsilon^{(2)}| = 2.23\,\varepsilon_0$. For the excited subbands, we find
$a_{1,-1/2,1}=-a_{-1,1/2,1}=0.693$ and $b_{1,-1/2,1}=b_{-1,1/2,1}= 0.436$.

The wave functions for the lowest subbands and first excited subbands are thus given by
\begin{subequations}
\begin{eqnarray}\label{eq:BoundWaveFun1}
\Psi^{(0)}_{(1,\frac{1}{2},1)}(r,\varphi)&=&\ee^{i(\frac{1}{2}-J_z)\varphi}~\Phi_{1,\frac{1}{2},1}^{(0)}(r)~,\\
\Psi^{(0)}_{(-1,-\frac{1}{2},1)}(r,\varphi)&=&\ee^{-i(\frac{1}{2}+J_z)\varphi}~\Phi_{-1,-\frac{1}{2},1}^{(0)}(r)~,
\end{eqnarray}
\end{subequations}
and
\begin{subequations}
\begin{eqnarray}
\Psi^{(0)}_{(1,-\frac{1}{2},1)}(r,\varphi)&=&\ee^{-i(\frac{1}{2}+J_z)\varphi}~\Phi_{1,-\frac{1}{2},1}^{(0)}(r)~,\\
\Psi^{(0)}_{(-1,\frac{1}{2},1)}(r,\varphi)&=&\ee^{i(\frac{1}{2}-J_z)\varphi}~\Phi_{-1,\frac{1}{2},1}^{(0)}(r)~,
\label{eq:BoundWaveFun2}
\end{eqnarray}
\end{subequations}
respectively.

The effective Hamiltonian of the cylindrical hole quantum wire for $k_z\neq0$ is obtained
by projection in the subspace spanned by states from
Eqs.~(\ref{eq:BoundWaveFun1})-(\ref{eq:BoundWaveFun2}):
\begin{eqnarray}\label{eq:EffectiveHam}
\langle \Psi^{(0)}_{(\sigma',\nu,n')}|\ham_{\text{v}}^{(\text{eff})}|\Psi^{(0)}_{(\sigma,\nu,n)}\rangle~,
\end{eqnarray}
where the inner product is $\langle\cdots\rangle=\int_0^R dr r \int_0^{2\pi}d\varphi\cdots$. 
The matrix of the superconducting pair potential can be obtained straightforwardly
by first transforming the matrix in Eq.~(\ref{eq:Gapmatrix}) into cylindrical coordinates
and then calculating the matrix elements with respect to the basis states in
Eqs.~(\ref{eq:BoundWaveFun1})-(\ref{eq:BoundWaveFun2}).

%

\end{document}